\documentclass[10pt]{wlscirep}
\usepackage[utf8]{inputenc}
\usepackage[T1]{fontenc}
\usepackage[english]{babel}
\usepackage{graphicx}
\usepackage{bm}
\usepackage{amsmath}
\usepackage{amsthm}
\usepackage{bbold}
\usepackage{soul}
\usepackage{filecontents}
\usepackage{tikz-cd}
\usepackage{float}
\usepackage{placeins}
\usepackage{amsthm}
\providecommand{\moy}[1]{\langle #1 \rangle}

\def\aop{\hat{a}}
\def\adag{\hat{a}^{\dagger}}
\def\Jx{\hat{J}_x}
\def\Jy{\hat{J}_y}
\def\Jz{\hat{J}_z}
\def\Xop{\hat{X}}
\def\Yop{\hat{Y}}
\def\Gamd{\Gamma_\downarrow}
\def\Gamp{\Gamma_\phi}
\def\Gp{\Gamma^\prime}

\begin{document}
\title{Dissipation-induced bistability\\ in the two-photon Dicke model}

\author{Louis Garbe$^{1,2}$}
\author{Peregrine Wade$^{1}$}
\author{Fabrizio Minganti$^{2}$}
\author{Nathan Shammah$^{2}$}
\author{Simone Felicetti$^{3}$}
\author{Franco Nori$^{2,4}$}
\affil{$^{1}$Laboratoire Mat\' eriaux et Ph\' enom\`enes Quantiques, Sorbonne Paris Cit\' e, Universit\' e de Paris, CNRS UMR 7162, 75013, Paris, France}
\affil{$^{2}$Theoretical Quantum Physics Laboratory, RIKEN Cluster for Pioneering Research, Wako-shi, Saitama 351-0198, Japan}
\affil{$^{3}$Departamento de F\'isica Te\'orica de la Materia Condensada and Condensed Matter Physics Center (IFIMAC), Universidad Aut\'onoma de Madrid, E- 28049 Madrid, Spain}
\affil{$^{4}$Physics Department, The University of Michigan, Ann Arbor, Michigan, 48109-1040, USA}

\begin{abstract}
The Dicke model\cite{Dicke54} is a paradigmatic quantum-optical model describing the interaction of a collection of two-level systems with a single bosonic mode. Effective implementations of this model made it possible to observe the emergence of superradiance, i.e., cooperative phenomena arising from the collective nature of light-matter interactions. Via reservoir engineering and analogue quantum simulation techniques, current experimental platforms allow us not only to implement the Dicke model, but also to design more exotic interactions, such as the two-photon Dicke model. In the Hamiltonian case, this model presents an interesting phase diagram characterized by two quantum criticalities: a superradiant phase transition and a spectral collapse, that is, the coalescence of discrete energy levels into a continuous band. Here, we investigate the effects of both qubit and photon dissipation on the phase transition and on the instability induced by the spectral collapse. Using a mean-field decoupling approximation, we analytically obtain the steady-state expectation values of the observables signaling a symmetry breaking, identifying a first-order phase transition from the normal to the superradiant phase. Our stability analysis unveils a very rich phase diagram, which features stable, bistable, and unstable phases depending on dissipation rate. 
\end{abstract}

\vskip2pc 
 
\vskip2pc 
\maketitle

The Dicke model describes the interaction of a collection of two-level systems with a single bosonic mode. In the thermodynamic limit, this model is known to exhibit a superradiant phase transition at zero temperature \cite{Dicke54,Hepp73,emary_chaos_2003,Lambert04,brandes_coherent_2005,Sachdev_2007,Kirton_2019,lambert_quantum_2009}. 
Namely, the ground-state number of photons changes non analytically form zero to  finite values as the light-matter coupling strength is increased across its critical value.
The capability of the Dicke model of capturing the physics of light-matter coupling near the critical point is the object of an ongoing debate.
In particular, it is unclear whether a \textit{genuine} light-matter coupling can give rise to a superradiant phase transition, due to the presence of the so-called diamagnetic term \cite{Nataf10,de_liberato_light-matter_2014,vukics_elimination_2014,de_bernardis_breakdown_2018, stefano_resolution_2019,Andolina_2019}. It is possible, however, to circumvent this controversy entirely by using driven systems and bath engineering to simulate effective Hamiltonians. This made it possible to observe the superradiant phase transition in driven atomic systems \cite{baden_realization_2014,klinder_dynamical_2015}.
In general, the implementation of analogue quantum simulations \cite{buluta_quantum_2009,georgescu_quantum_2014} provide an ideal playground to test driven-dissipative physics in a controlled setting. Their experimental feasibility motivates increasing research efforts devoted to the study of driven-dissipative quantum optical models, as it is known that noise and dissipation can drastically change the properties of the steady-state phase diagrams and the emergence of phase transitions \cite{MingantPRA18,KesslerPRA12}.
Analytical studies\cite{keldysh_dicke_2013,Sieberer_2016,kirton_suppressing_2017,hwang_dissipative_2018} are extremely challenging, as the intrinsic non-equilibrium nature of driven-dissipative systems does not allow a determination of the stationary state of the system via a free energy analysis \cite{Carmichael_BOOK_1,zoller_2004,Haroche_BOOK_Quantum,BreuerBookOpen}.

Effective implementations offer others significant advantages. In a cavity-QED setting, driving the system allows to turn virtual excitations inside the cavity into real excitations which can exit the cavity\cite{MarcovicPRL18}, thus giving an immediate access to the intra-cavity dynamics.
By manipulating the exchanges between a system and its environment, reservoir-engineering techniques allow us to realise previously inaccessible quantum phases of matter ~\cite{PoyatosPRL96,VerstraeteNATPH2009,TanPRA13,ArenzJPB13,AsjadJPB14,RoyPRA15,LeghtasScience15}.
For instance, it is possible to stabilize phases without an equilibrium counterpart \cite{LeePRA11,LeePRL13,JinPRX16}, and reservoir engineering methods for complex many-body phases have been thoroughly explored in different contexts \cite{DiehlNATPH2008,VerstraeteNATPH2009}. Several experiments and theoretical proposals have applied these ideas to study generalized Dicke models \cite{baden_realization_2014,zou_implementation_2014,genway_generalized_2014,klinder_dynamical_2015,felicetti_spectral_2015,felicetti_two-photon_2018,felicetti_ultrastrong-coupling_2018} and the ultrastrong coupling regime (USC) \cite{Kockum19,Forn19,Gunter09,Schwartz11,Porer12,Scalari12,Garziano13,Stassi16,DeLiberato17,Qin18,Leroux18,Cirio16,Cirio19}, i.e., the regime of parameters where the coupling constant becomes a sizable fraction of both the qubit and bosonic frequencies.

Furthermore, driving and bath engineering can be exploited to control not only the dissipation of the system, but also the form of the coherent light-matter interaction. Among the variety of phenomena which are made accessible by analogue quantum simulations \cite{buluta_quantum_2009,georgescu_quantum_2014}, a particularly interesting one is the possibility to engineer a coupling involving the simultaneous exchange of several photons. In superconducting circuits\cite{gu_microwave_2017}, this is possible with nonlinear virtual processes in the USC regime \cite{Garziano15,Kockum17,Kockum17b}, and experimentally a photon-pair driving mechanism has been realized~\cite{LeghtasScience15},
leading to the generation of so-called photonic Schr\"odinger cats~\cite{KerckhoffOE11,MingantiSciRep16,SavonaPRA17}.
The possibility to control and protect such states is promising for the implementation of quantum computation protocols~\cite{GilchristJOB04,OurjoumtsevScience06,MirrahimiNJP14,GotoPRA16,PurinpjQI17}.
In the two-photon Dicke model \cite{felicetti_spectral_2015,garbe_superradiant_2017,felicetti_two-photon_2018}, it is the light-matter interaction that creates or annihilates one pair of bosonic excitations per qubit exchange.
This exotic exchange leads to several unusual properties in the USC regime \cite{Kockum19,Forn19}. 
In particular, there exists a critical value of the coupling strength where the discrete spectrum collapses into a continuous band \cite{felicetti_spectral_2015}. 
For higher values of the coupling, the two-photon Dicke Hamiltonian is no longer bounded from below, indicating the breaking down of the model itself. 
It was recently shown that, in spite of the spectral collapse, a superradiant-like phase transition can take place also in the two-photon Dicke model \cite{garbe_superradiant_2017}. This transition has been characterized also for other two-photon interaction models \cite{ChenPRA18,CuiPRA19} but, so far, only the Hamiltonian case has been considered.

In this paper, we present a theoretical analysis of a two-photon interaction model in the driven-dissipative case.
In particular, we consider the $N$-body two-photon Dicke model connected to an engineered Markovian bath. The Lindblad formalism is used to introduce different incoherent processes, such as photon loss $\kappa$, individual qubit decay $\Gamma_\downarrow$, and local qubit dephasing $\Gamma_\phi$. 
To analyze the system dynamical properties, we resort to a mean-field decoupling of the equation of motion, allowing to determine (semi-)analytically the steady-state of the system. We show the emergence of a first-order phase transition from a normal to a superradiant phase, for a critical value of the light-matter coupling $g$. A numerical study of the stability of the different phases unveils a very rich phase diagram: depending on the strength of the atomic dissipation, the system may be stable, bistable, or unstable. Interestingly, we find that atomic dephasing can be beneficial to the stabilization of the superradiant phase, a conceptual difference  with respect to the phenomenology of the standard Dicke model \cite{kirton_suppressing_2017}.

This paper is organized as follows: first, we introduce the one- and two-photon Dicke model, and briefly describe some of their properties using heuristic arguments. 
Secondly, we discuss dissipation in the two-photon model. Thirdly, we present the phase diagram of the system obtained via a decoupling mean-field approximation, and show the existence of two different regimes of  dissipation. Finally, we summarize our conclusions and present some perspectives for future work.

\section*{One-and two-photon Dicke models}

The standard Dicke model was originally used to describe the behaviour of a collection of atoms with the electromagnetic (EM) field inside a high-quality-factor cavity. It can be derived taking  several assumptions about the system. For instance, the atomic size must be small with respect to the field wavelength, making the atoms insensitive to the field modulation \cite{Dicke54,Haroche_BOOK_Quantum,Shammah17}. The atoms must couple to a single mode of the field. Finally, the atomic energy level structure must be highly anharmonic, so that only one transition is resonant with the field, allowing us to approximate the atoms by two-level systems (or qubits). This so-called two-level approximation, when handled improperly, can lead to a gauge ambiguity in the USC regime that has been fully resolved only recently \cite{stefano_resolution_2019}. When these assumptions hold, the system is described by the one-photon Dicke Hamiltonian (here $\hbar=1$),

\begin{equation}
\label{onephotonDicke}
	\hat{H}_1=\omega_c\adag \aop+\omega_0\sum^N_{j=1}\hat{\sigma}^j_z+ \frac{g}{\sqrt{N}}\sum^N_{j=1}\hat{\sigma}^j_x\left(\aop+\adag\right),
\end{equation}

where $\aop$ ($\adag$) is the annihilation (creation) operator of the bosonic mode, $N$ is the number of qubits, and $\sigma_{x, y, z}^j$ are the Pauli matrices describing the $j$-th qubit. This Hamiltonian exhibits a $\mathbb{Z}_2$ symmetry, corresponding to the simultaneous exchange
\begin{equation}\label{Eq:Symmetry-transform_one}
\left\lbrace
\begin{split}
&\aop \to -\aop, \\
&\hat{\sigma}_x^j\rightarrow -\hat{\sigma}_x^j \quad \forall j.
\end{split}
\right.
\end{equation}

For low values of the coupling, the ground state of this Hamiltonian is given by the product state of the field vacuum and the atoms individual ground states. When the coupling constant enters the USC regime, however, the system experiences a second-order phase transition in the thermodynamic limit which breaks the $\mathbb{Z}_2$ symmetry. The system enters the so-called superradiant phase, in which the bosonic field is described by a coherent state, while the qubits are rotated by a common angle \cite{emary_chaos_2003}.
The possibilities offered by quantum simulations have brought this model far beyond the atom-cavity setting. For instance, the cavity EM field may be replaced by microwave resonators in superconducting circuits, or by vibrational motion in atomic platforms.  These effective implementation made it possible to circumvent the problems raised by gauge ambiguities and to observe the superradiant phase transition\cite{baden_realization_2014,klinder_dynamical_2015}.

These platforms also pave the way to the experimental exploration of novel forms of light-matter interactions. In particular, quantum simulation schemes make it possible to implement two-photon interaction models both in the SC and in the USC regime \cite{Hayn11,felicetti_spectral_2015,puebla_probing_2017,ChengPRA18}. For instance, in trapped-ions experiments, laser-induced interactions can be used to couple the internal state of the ions to their motional degrees of freedom.  Let us assume that the properties of the trap allows to single out a single vibrational mode with frequency $\nu$. If the detuning between the laser and the internal transition is close to $-2\nu$ (red-detuned laser), then the laser can excite a process in which two phonons are destroyed and one qubit excitation is created. If the detuning is close to $2\nu$ (blue-detuned laser), then the energy brought by the laser can be used to simultaneously create one qubit excitation and two phonons. Therefore, by using both a red-detuned and a blue-detuned lasers, one can engineer a qubit-boson coupling similar to the Dicke model, but where the standard one-boson interaction term is replaced by a two-boson term, which is generically called in the litterature two-photon or two-phonon coupling term. For simplicity, we will only use the term "two-photon" and the cavity-QED terminology in the following. Furthermore, the modulation of photonic states by the laser pump permits to effectively renormalize both the bosonic frequency and the coupling constant, thus allowing to bring the two-photon coupling to the USC regime.

Similarly, it has recently been shown \cite{felicetti_two-photon_2018,felicetti_ultrastrong-coupling_2018} that two-photon interactions can also be implemented in superconducting circuits, engineering an intrinsic nondipolar coupling between a superconducting artificial atom and superconducting quantum interference device (SQUID). In this case, the standard linear coupling is suppressed, while the two-photon coupling terms emerge as the natural light-matter interaction in an undriven system and  not as the result of a quantum simulation scheme.

The possibility of implementing the two-photon coupling term motivates the study of the \textit{two-photon} Dicke model, whose  Hamiltonian reads (setting $\hbar=1$), 
\begin{equation}
\label{two-photonDicke}
\hat{H}=\omega_c\adag \aop+\omega_0\sum^N_{j=1}\hat{\sigma}^j_z+ \frac{g}{\sqrt{N}}\sum^N_{j=1}\hat{\sigma}^j_x\left[\aop^2+\left(\adag\right)^2\right].
\end{equation}
This Hamiltonian exhibits a four-folded symmetry, stemming from the simultaneous exchange of
\begin{equation}\label{Eq:Symmetry-transform}
\left\lbrace
\begin{split}
&\aop \to i\aop, \\
&\hat{\sigma}_x^j\rightarrow -\hat{\sigma}_x^j \quad \forall j.
\end{split}
\right.
\end{equation}

In the USC limit, this model exhibits \cite{felicetti_spectral_2015} an instability known as spectral collapse, where the discrete spectrum collapses into a continuous band for a critical value of $g$. Some intuition about this effect can be gained through the following reasoning. When the coupling constant $g$ in Eq.~\eqref{two-photonDicke} becomes large, the interaction term dominates the physics. Since this term commutes with the $\hat{\sigma}^i_x$, we can study the qubits domains $\hat{\sigma}^i_x=-\frac{1}{2}$ and $\hat{\sigma}^i_x=+\frac{1}{2}$ independently. Let us consider $\hat{\sigma}^i_x=-\frac{1}{2}$ for all $i$. Then we have an effective boson dynamics described by this Hamiltonian
\begin{equation}
\omega_c\adag\aop-\frac{g\sqrt{N}}{2}\left[\aop^2+(\adag)^2\right]=\left(\frac{\omega_c}{4}-\frac{g\sqrt{N}}{4}\right)\hat{x}^2+\left(\frac{\omega_c}{4}+\frac{g\sqrt{N}}{4}\right)\hat{p}^2, 
\end{equation}
which is a quadratic potential for the field quadratures $\hat{x}=\adag+\aop$ and $\hat{p}=i(\adag-\aop)$. When $g$ is large enough, this potential becomes almost flat, shrinking the gap between the different energy levels. Ultimately, these levels coalesce into a continuous band, causing the so-called spectral collapse \cite{felicetti_spectral_2015}. When $g$ is increased even further, the potential becomes an upside-down harmonic well, and is unbounded from below for $\hat{x}\rightarrow\infty$.
Therefore, the dynamics of the system will become unstable, signaling the breaking down of the model. By contrast, in the one-photon Dicke model \eqref{onephotonDicke}, the interaction term adds only a linear correction, meaning that the Hamiltonian can never be unbounded from below.

Very recently it has been shown\cite{garbe_superradiant_2017} that the two-photon Dicke model can also display a second-order quantum phase transition very similar to the superradiant transition of the one-photon Dicke model. Instead of a coherent state, however, the bosonic field here will be described by a squeezed state for high values of the coupling. The ground-state phase diagram has been analyzed with different numerical and analytical techniques also for other two-photon coupling models\cite{ChenPRA18,CuiPRA19}. However, two-photon light-matter interaction models have so far never been considered from an open-quantum-system perspective.

\section*{Effect of dissipation}

The physics of the one-photon Dicke model changes drastically once dissipation is taken into account. It was shown in Refs.~\cite{DallaTorre_PRA2016,kirton_suppressing_2017} that in the presence of qubit decay and dephasing, the transition of the Dicke model could be modified, supressed, or restored. Similarly, the presence of dissipative processes in the two-photon Dicke model raises intriguing questions. 

Assuming a Markovian environment and performing the Born approximation \cite{Haroche_BOOK_Quantum}, the dissipation may be described by a Lindblad master equation\cite{zoller_2004,ulrich_2012}.
In our anaysis, we will assume that the Hamiltonian part of the evolution remains that of Eq.~\eqref{two-photonDicke}, and we will include three dissipation channels, that is, individual qubit decay and dephasing, and photon loss. We obtain the following Lindblad equation (we recall $\hbar=1$),

\begin{equation}
\label{Lindblad}
\partial_t\hat{\rho}=-i[\hat{H},\rho]+\kappa \mathcal D[\aop](\hat{\rho}) + \sum_{j=1}^N \Gamma_{\downarrow}\mathcal D[\hat{\sigma}^j_-](\hat{\rho}) + \Gamma_{\phi}\mathcal D[\hat{\sigma}^j_z](\hat{\rho}), 
\end{equation}
where $\hat{\rho}(t)$ is the density matrix of the system at time $t$, $\hat{\sigma}_-^j=\hat{\sigma}_x^j - i\hat{\sigma}_y^j$, $[\hat{H},\hat{\rho}]$ indicates the commutator, and $\mathcal D[\hat{A}]$ are the Lindblad dissipation superoperators defined as \cite{zoller_2004,Haroche_BOOK_Quantum,Carmichael_BOOK_1}
\begin{equation}
\mathcal D[\hat{A}]\hat{\rho}=\hat{A}\hat{\rho}\hat{A}^{\dagger}-\frac{1}{2}\hat{\rho}\hat{A}^{\dagger}\hat{A}-\frac{1}{2}\hat{A}^{\dagger}\hat{A}\hat{\rho}.
\end{equation}
Even if the Hamiltonian symmetry cannot be directly translated into one for the Lindblad master equation \cite{albert_symmetries_2014}, Eq.~\eqref{Lindblad} remains identical upon the transformation in Eq.~\eqref{Eq:Symmetry-transform}.
In this regard, also the Lindblad master equation presents a four-folded symmetry similar to that of the Hamiltonian case.
Let us note that this equation is a purely phenomenological one and, while it is the most appropriate in a quantum simulation framework, for a genuine implementation of the model in the USC regime it would fail to describe the true evolution. Indeed, in the presence of bare local dephasing and qubit decay processes, the system would not tend towards the dressed ground state. In fact, these processes would effectively pump energy into the system, forcing it away from the true polaritonic ground state and toward a different steady state; while considering dressed-operator incoherent processes would lead to the polaritonic ground state \cite{shammah_open_2018}. Moreover, a microscopic theory needs to be developed for arbitrary strengths of the dissipative couplings, which has shown that USC effects can be robust in loss-dominated systems \cite{DeLiberato17}.  
However, our analysis is meant to describe effective implementations of the model, where the considered decoherence and dissipation processes can themselves be implemented via bath-engineering techniques \cite{puebla_probing_2017}. 
For instance, this Lindbladian dynamics could be observed in a strongly driven atomic cloud, where Raman processes effectively engineer the wanted processes \cite{Dimer07,Baumann10}. Note also that, since quantum simulation allows to renormalize both the effective frequencies $\omega_c$ and $\omega_0$ and the coupling constant $g$ \cite{felicetti_spectral_2015}, the dissipation constant may  be large with respect to $\omega_c$, $\omega_0$ and $g$, while remaining small with respect to the actual frequencies of the system.

We expect that this model will have very different behavior from its Hamiltonian counterpart. One the one hand, the critical behaviour and the properties of the superradiant phase can drastically change, as for the one-photon model \cite{kirton_suppressing_2017,Kirton_2019}. On the other hand, the spectral collapse may be modified or avoided due to the presence of dissipation. Indeed, the photon loss term in Eq.~\eqref{Lindblad} acts like a stabilizing quadratic term which can balance the effect of the Hamiltonian unstable potential.

\section*{Results}
\subsection*{Symmetry breaking}
We have focused our analysis on the following quantities: $\hat{J}_{u=x,y,z}=\frac{1}{N}\sum_{j=1}^N\hat{\sigma}_u^{j}$, $\Xop=\aop^2+\hat{a}^{\dagger2}$, $\Yop=\aop^2-\hat{a}^{\dagger2}$, $\adag\aop$. 
The Hamiltonian treatment \cite{garbe_superradiant_2017} predicts a symmetry breaking during which the bosonic field becomes squeezed, which is captured by the second-order moment of the bosonic field.
Therefore, we expect that the observables $\Xop$, $\Yop$ and $\adag\aop$ will be valid order parameters also in the presence of dissipation. 
We have studied the evolution of these quantities using a mean-field decoupling approximation (see the Methods section). We found that the dynamics of these quantities has three possible solutions. The first one corresponds to $\moy{\adag\aop}=\moy{\Xop}=\moy{\Yop}=\moy{\Jx}=\moy{\Jy}=0$ and $\moy{\Jz}=-1$. The other two phases have $\moy{\adag\aop}$, $\moy{\Xop}=\pm X_s$, $\moy{\Yop}=\pm Y_s\neq0$ (complete expressions in the Methods section, Eq. \eqref{Eq:solution}). In accordance with previous results in the one- and two-photon Dicke model\cite{Kirton_2019,emary_chaos_2003,kirton_suppressing_2017,garbe_superradiant_2017}, we can identify the first solution as the "normal phase", as it corresponds to the product state of the individual ground states of the field and the atoms. The other two solutions correspond to the "superradiant phase"; that is, they contain a macroscopic number of atomic and photonic excitations. Since the "superradiant" solutions have $\moy{\Xop}\neq0$, the four-folded symmetry of the model is at least partially broken. The stability of each phase has been analyzed numerically (see Methods). Let us now illustrate the properties of the superradiant phase and the complex driven-dissipative phase diagram of the model considered.

\subsection*{Nature of the phase transition}

\begin{figure}[ht]
\begin{center}
	\includegraphics[width=0.55\linewidth]{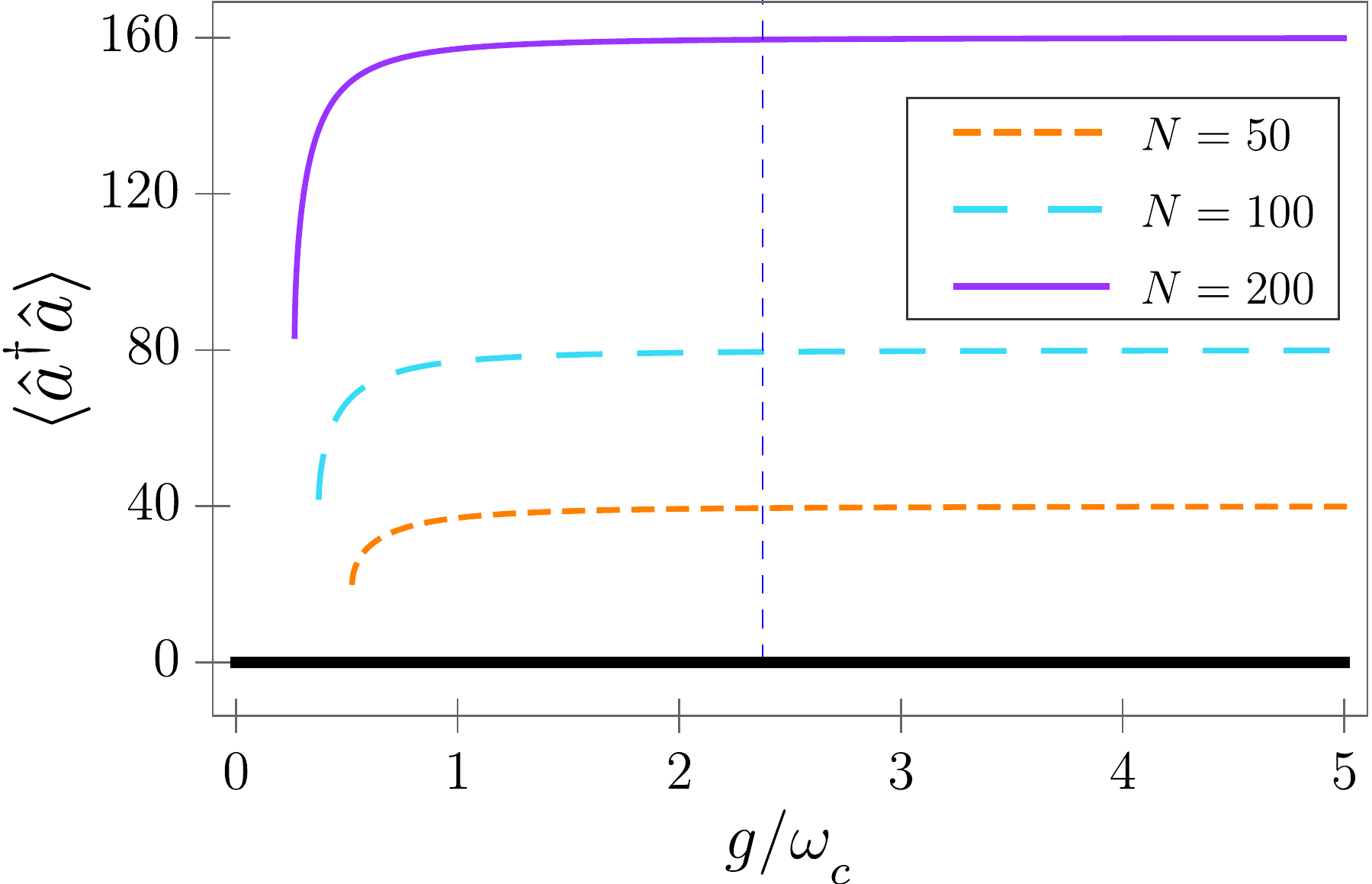}
\end{center}	
	\caption{Photon number in the superradiant phase versus normalised light-matter coupling, for several qubit numbers $N$. We have set $\kappa=\omega_c$ and $\Gamd=\Gamp=3\omega_c$. For small $g$, the superradiant phase becomes unphysical. The horizontal black line is a guide to the eye indicating the value $\langle\adag\aop\rangle=0$, which is the value the field adopts in the normal phase. The vertical dashed line indicates the point where the normal phase becomes unstable (this point is independent of $N$). In all cases, the superradiant phase becomes stable before the normal phase becomes unstable, indicating bistability.}
	\label{photonb}
\end{figure}

In Figure~\ref{photonb} we show the value of the steady-state photon number in the superradiant phase as a function of the coupling strength $g$, $\omega_0=\omega_c$, and $\kappa=\omega_c$. For now, we have set $\Gamd=\Gamp=\Gamma$, and $\Gamma=3\omega_c$. For small values of $g$, the superradiant phase yields nonphysical complex values for $\langle \adag \aop\rangle$, showing that the system can only reach the normal phase $\langle \adag \aop\rangle=0$ until the critical value of the coupling strength is achieved. When $g$ is increased, the superradiant phase becomes physical, and the stability analysis reveals it is stable as well. 
Therefore, a nonzero number of bosonic excitations can appear in the system. Analyzing the average photon number in the system steady state, we can already identify two qualitative differences with respect to the ground state in the Hamiltonian case. First, in the driven-dissipative case the number of photons in the superradiant phase does not go to zero when one approaches the limit of stability from above. Second, the point at which the normal phase becomes unstable and the superradiant phase becomes stable do not coincide. Therefore, the driven-dissipative two-photon Dicke model exhibits bistability at the mean-field level.

The emergence of bistability in mean-field models is well-known in open quantum systems. 
A typical example is that of the Kerr resonator, where the semiclassical solution obtained via the Gross-Pitaevskii mean field has three different solutions: two which are stable and one unstable.  As soon as one considers the quantum steady state, however, only one solution is found \cite{Drummond_JPA_80_bistability}. This apparent contradiction can be solved by considering the full Liouvillian spectrum, where the onset of bistability is in close relation to the emergence of criticality \cite{BartoloPRA16,MingantPRA18,Landa19}. 
Indeed, several models presenting bistable behaviour at the mean-field level proved to display a genuine first-order phase transition in the thermodynamic limit of a full quantum model
\cite{BiondiPRA17,BiellaPRA17,SavonaPRA17,JinPRB18,Foss-FeigPRA17,VicentiniPRA18,RotaPRL19,LeBoitePRL13,LeBoitePRA14}.

These results show that the Hamiltonian and dissipative versions of this model are strikingly different. In the equilibrium case, a second-order phase transition is predicted to occur, and only in the far-detuned regime \cite{garbe_superradiant_2017} $\omega_0\ll\omega_c$. In the nonequilibrium case, a first-order phase transition takes place in the resonant regime $\omega_0=2\omega_c$, a condition that strongly simplifies possible experimental implementations. 

\subsection*{Phase diagram}
\begin{figure}[ht]
\includegraphics[width=\linewidth]{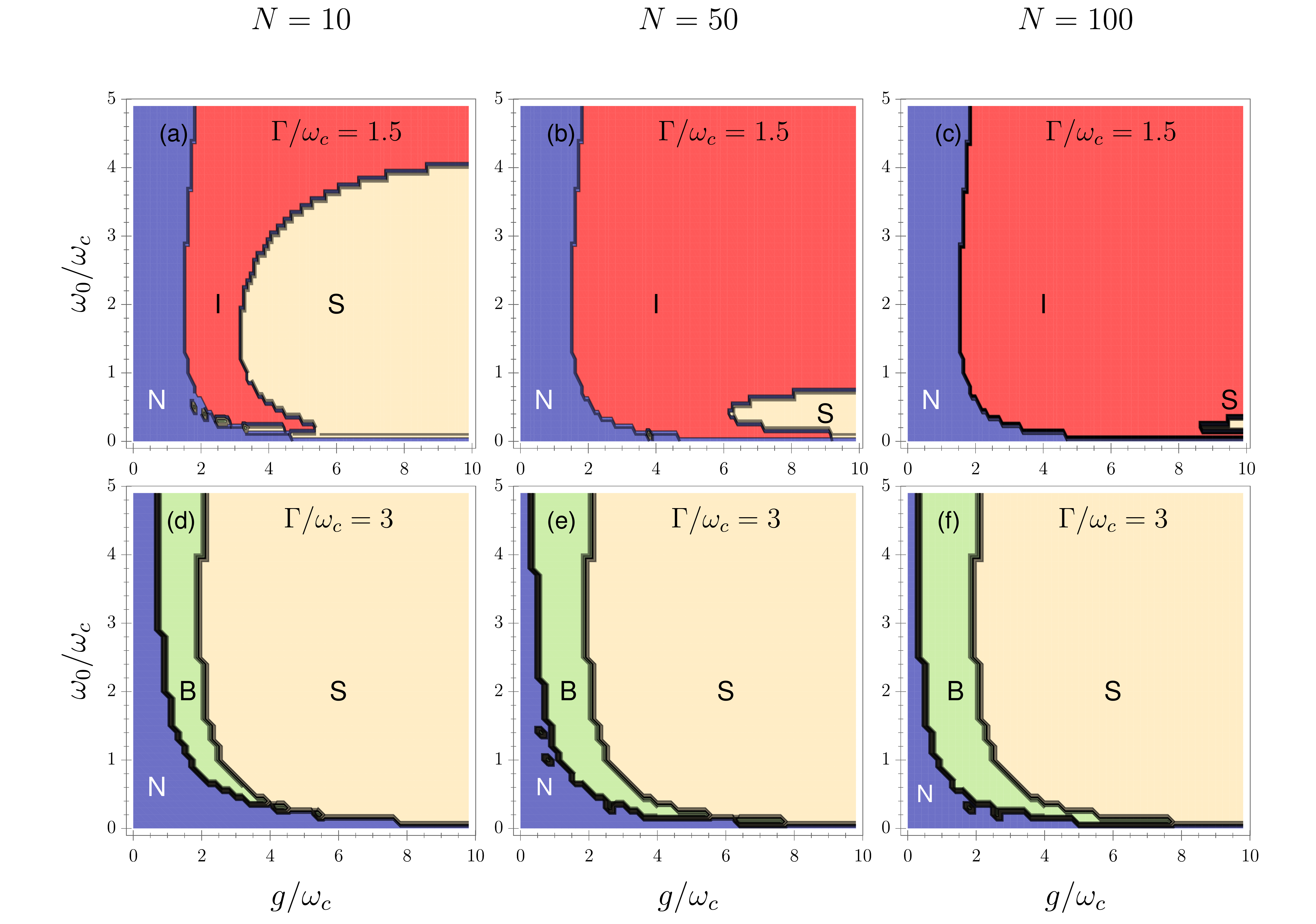} 
\caption{Phase diagram of the system for various $\Gamma$ and $N$, with $\kappa=\omega_c$. Upper row: $\Gamma/\omega_c=1.5$, lower row $\Gamma/\omega_c=3$. N: normal phase, S: superradiant phase, B: bistable phase, I: instability. For $\Gamma/\omega_c=1.5$, the region of stability for the superradiant phase shrinks when the number of qubits increases. For $\Gamma/\omega_c=3$, a bistable behavior is observed, and the phase diagram is almost invariant when the number of qubits is increased beyond a few dozens.}
\label{gw0}
\end{figure}

Having established the existence of a phase transition, we can produce the phase diagram of the model by studying the stability of both phases for a broad range of parameters. 
The analysis of these diagrams revealed the existence of two regimes of dissipation.
In Fig.~\ref{gw0}, we display the phase diagram in the $g$-$\omega_0$ plane, for two values of  $\Gamma$: $\Gamma=1.5\omega_c$ and $\Gamma=3\omega_c$, and for various number of qubits $N$. For $\Gamma=1.5\omega_c$ and the smaller value $N=10$ qubits, we observe that the mean-field equations predict the existence of a zone where the superradiant phase is stable. However, the size of this zone shrinks when $N$ increases. Since the mean-field description becomes correct only for $N\rightarrow\infty$, \textit{no phase transition can happen in the mean field limit for this value of dissipation}. The system will either reach the normal steady-state or be unstable. 
For $\Gamma=3\omega_c$, however, we observe that bistability become possible. In the thermodynamic limit, the region of stability becomes independent of the number of qubits, meaning that \textit{a phase transition can take place in the $N\to\infty$ limit}. 
Values of $\Gamma/\omega_c$ lower that $1.5$ or higher than $3$ yield qualitatively similar results, which allows us to conclude that there are two regimes of dissipation: a large dissipation regime in which a phase transition is possible, and a low dissipation regime in which only the normal phase is stable in the thermodynamic limit. 

Interestingly, the transition between these two regimes of parameters when $\Gamma$ is increased is quite sharp, especially in the thermodynamic limit. To visualize this, we study the stability of the superradiant phase versus both $g$ and $\Gamma$, for 100 qubits, and for various values of $\omega_0$, the other parameters being the same (this amounts to taking horizontal slices in Figure~\ref{gw0} and study their evolution when $\Gamma$ changes).

\begin{figure}[ht]
\includegraphics[ width=\linewidth]{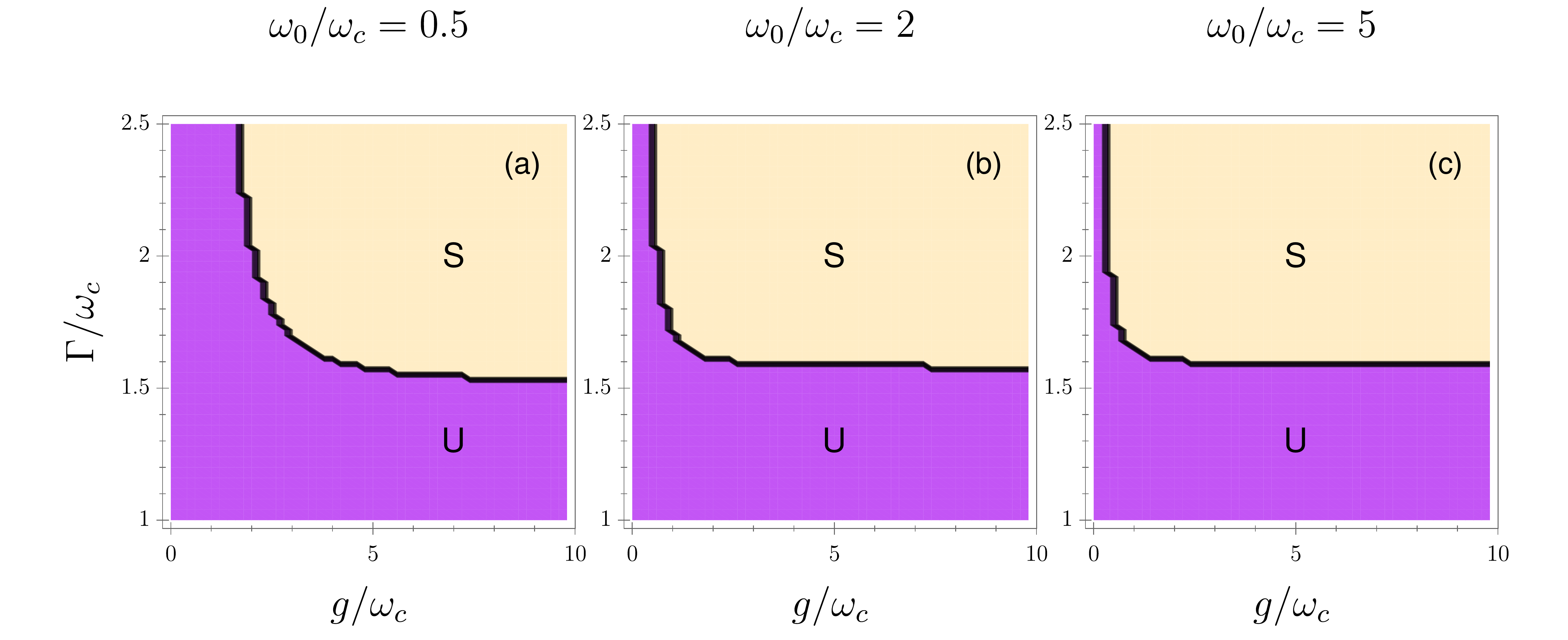}
\caption{Stability of the superradiant phase versus coupling and dissipation, for $\kappa=\omega_c$ and $100$ qubits. Left: $\omega_0/\omega_c=0.5$, Center: $\omega_0/\omega_c=2$, Right: $\omega_0/\omega_c=5$. S: stable superradiant phase, U: unstable superradiant phase. Within the U phase, the normal phase may be stable or unstable, but for simplicity we have chosen not to represent it. Except for small values of $\omega_0/\omega_c$ or $g$, the value of $\Gamma/\omega_c$ for which the transition occurs is almost the same for all parameters, around $1.6$. This means the phase diagram as a whole changes drastically when $\Gamma$ goes across this value.}
\label{gvsGamma}
\end{figure}

The results are displayed in Fig.~\ref{gvsGamma}: for $\Gamma/\omega_c\approx 1.6$, the instability disappears and the superradiant phase becomes stable for most values of $\omega_0$ and $g$. Hence, the phase diagram as a whole changes drastically when $\Gamma/\omega_c$ goes across this threshold.

\begin{figure}[ht]
    \centering
    \includegraphics[width=.9\linewidth]{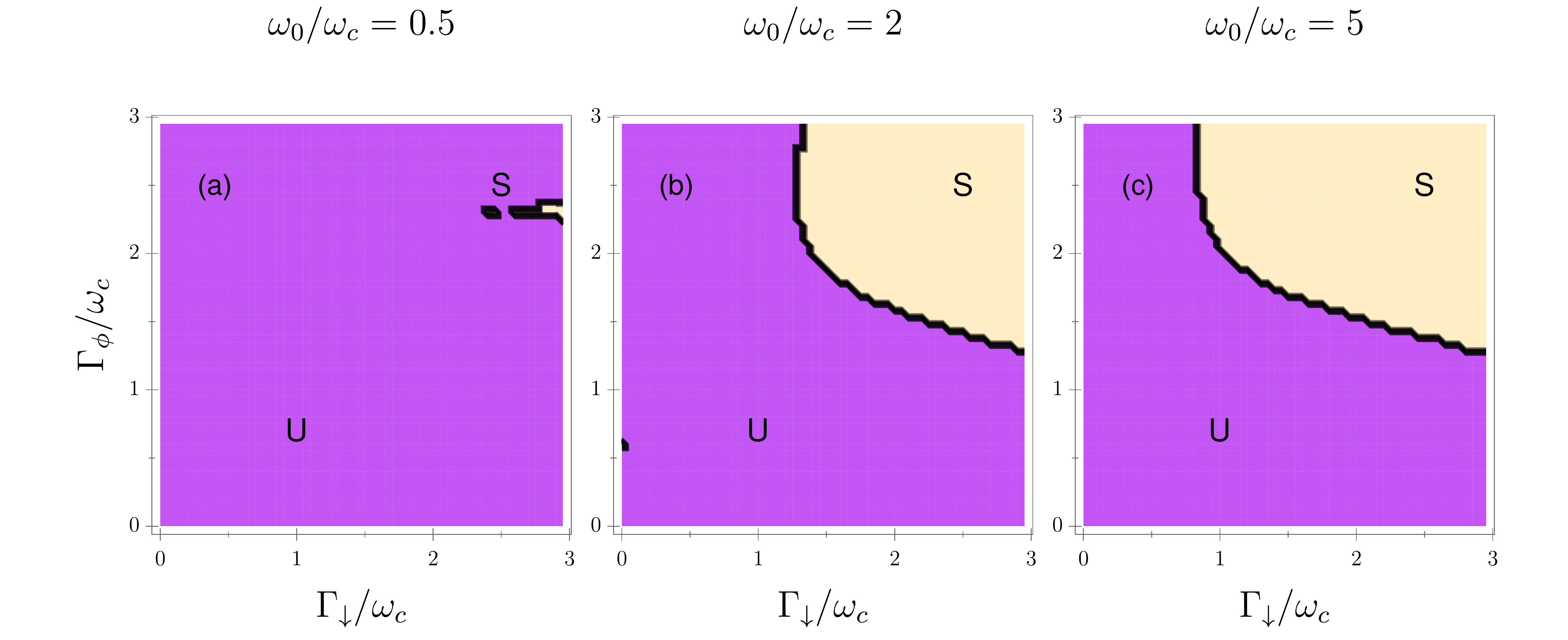}
    \caption{Stability of the superradiant phase versus qubit dephasing and decay. Here $\kappa/\omega_c=1$, $g/\omega_c=1$, and $N=100$ qubits. Left: $\omega_0/\omega_c=0.5$, center: $\omega_0/\omega_c=2$, right: $\omega_0/\omega_c=5$. U: unstable superradiant phase, S: stable superradiant phase. Here, both dephasing and decay contribute positively to the stabilization of the superradiant phase.}
    \label{gGammaGamma}
\end{figure}

Hence, we have established that the presence of dissipation is instrumental in stabilizing the superradiant phase. If we compare this with the results obtained in the one-photon version of the driven-dissipative model \cite{kirton_suppressing_2017}, an instructive analogy can be made. Adding enough qubit dissipation appears to preserve the superradiant phase transition, which normally would be destroyed in the presence of noise. In the one-photon case, however, decay and dephasing play antagonistic roles: adding an infinitesimal amount of qubit dephasing destroys the transition, while qubit decay stabilizes it. To see if such effect is also present in the two-photon model, we study the stability of the superradiant phase with respect to both $\Gamp$ and $\Gamd$. Results are displayed in Figure~\ref{gGammaGamma}. We see no evidence of supression and restoration of the phase transition. Rather, these plots indicate that both dephasing and decay contribute positively to the stabilisation of the superradiant phase.

{}

\section*{Discussion}
\label{Discussion}
In this paper, we present the first analysis of the steady-state phase diagram of a  two-photon interaction model in the driven-dissipative case. In particular, we have explored numerically the mean-field behavior of the $N$-body two-photon Dicke model. We have identified a rich behavior, including a superradiant phase transition of first order, a bistable phase, and an instability that is removed by dissipation. This instability may be linked with the spectral collapse at Hamiltonian level. The equivalence, though is not perfect. In the Hamiltonian case, the spectral collapse is expected to occur when $g$ increases. In our case, for $\Gamma=3\omega_c$, we have increased $g$ towards higher values, up to $10^3$ (not shown). We have found only a stable superradiant phase, even if, in the limit of very large $g$, we also expect the mean-field decoupling approximation to break down. 

These results indicate that this dissipative two-photon Dicke model is strikingly different from its Hamiltonian counterpart, which exhibits a second-order phase transition and a spectral collapse. This illustrates how the critical behavior of a given phase transition can change radically when one goes from the equilibrium case to the non-equilibrium one.
Qubit dissipation, such as local dephasing and decay, is clearly instrumental in controlling this behavior, and for sufficient dissipation the instability is expected to vanish, thus guaranteeing a phase transition at the mean-field level. 

Furthermore, we have pointed out conceptual differences between the behavior of the one- and two-photon Dicke models. For the latter, both local dephasing and decay appear to help stabilizing the transition, in contrast to what was found in the dissipative one-photon Dicke model. In the termodynamic limit of the two-photon Dicke model, the entire phase diagram changes very abruptly when dissipation is moved across a very narrow range of parameters.  

We note that while a mean-field approach can predict several solutions for the steady-state equation in the bistable region, and a unique solution in the mono-stable regions, in a finite-size quantum system there can only be a unique steady state in each region. Indeed, the introduction of both quantum and classical fluctuations prevents the fields from remaining stationary around their mean-field values. Several works \cite{BartoloPRA16,BartoloEPJST17} have illustrated these behaviours in the similar case of the two-photon Kerr resonator.
In turn, the presence of multiple solutions at the mean-field level is translated into an observable bistable behavior of the critical parameters of the full quantum model in a quantum trajectory approach \cite{MingantiPhD}.

Indeed for future perspectives, a full quantum treatment of the driven-dissipative two-photon Dicke model would be an interesting and yet challenging task. Extracting information on the thermodynamic limit from direct numerical simulations of the full dynamics is far from straightforward, due to the exponentially increasing Liouvillian space. Exploiting the permutational invariance of the Liouvillian \cite{shammah_open_2018} can exponentially reduce the computational overhead with regards to the qubit degrees of freedom, but the photonic subspace, approximated by a cut-off photon excitation number $n_\text{ph}$, needs to be larger than in the case of the single-photon Dicke model to avoid spurious results induced by the finite approximation of the otherwise unbounded Hilbert space. 
With this regard, two possible solutions could be considered, also simultaneously: first, the inclusion a two-photon dissipation process, which would effectively reduce the highest excitation number explored, for appropriately large values of the two-photon decay rate; second, the use of quantum trajectories, which reduces the computational overhead from being that of the Liouvillian space to just that of an effective Hilbert space, at the cost of averaging over many runs. We point out though that the intermittent dynamics characterizing a bi-stable phase can be grasped even by single quantum trajectory simulations. Alternatively to these approaches, a qubit-only description of the system could be obtained at the cost of abandoning the Lindbladian formalism for a full Redfield theory \cite{Damanet19}.

Finally, we have shown that in the driven-dissipative case the superradiant phase transition of the two-photon Dicke model can be observed also for resonant interactions, in contrast with the static case, where a large detuning is required. This result shows that this quantum phase transition can be observed in a more accessible parameter regime than previously thought. 
The driven-dissipative two-photon Dicke model could be implemented with trapped ions \cite{felicetti_spectral_2015} or in a cold atoms setup, similar to what has been done already for the driven Rabi model \cite{Schneeweiss_2018,Dareau_2018}. Superconducting circuits \cite{gu_microwave_2017} provide another platform to simulate this dynamics, in which the two-photon interaction can be engineered between a flux qubit and a SQUID resonator \cite{felicetti_two-photon_2018,felicetti_ultrastrong-coupling_2018}.

\section*{Methods}

From the Lindblad master equation~\eqref{Lindblad}, we can obtain the field equation governing the dynamics of any operator $\partial_t \moy{\hat{A}}$.
For the operators $\hat{X}$, $\hat{Y}$, $\adag\aop$, $\hat{J}_x$, $\hat{J}_y$, $\hat{J}_z$ that we have studied, this gives:
\begin{equation}\label{Eq:observables_not_closed}
    \begin{cases}
    \partial_t \moy{\hat{X}}=-\kappa \moy{\hat{X}}=-2i\omega_c \moy{\hat{Y}}, \\
    \partial_t \moy{\hat{Y}}=-\kappa \moy{\hat{Y}}-2i\omega_c \moy{\hat{X}} - 4ig\sqrt{N}\moy{\hat{J}_x} -8ig\sqrt{N}\moy{\hat{J}_x \adag\aop},\\ 
    \partial_t \moy{\adag\aop}=2i g \sqrt{N}\moy{\hat{J}_x\hat{Y}}-\kappa \moy{\adag\aop},\\
    \partial_t \moy{\hat{J}_x}=-2\omega_0\moy{\hat{J}_y} - \Gamma^\prime \moy{\hat{J}_x},\\
    \partial_t \moy{\hat{J}_y}=2\omega_0\moy{\hat{J}_x}-\Gamma^\prime \moy{\hat{J}_y}-\frac{2g}{\sqrt{N}}\moy{\hat{J}_z \hat{X}},\\
    \partial_t \moy{\hat{J}_z}=\frac{2g}{\sqrt{N}}\moy{\hat{J}_y \hat{X}}-\Gamma_{\downarrow}\moy{\hat{J}_z}-\Gamma_{\downarrow},
    \end{cases}
\end{equation}
where we have defined $\Gamma^\prime =2\Gamma_{\phi}+\frac{\Gamma_{\downarrow}}{2}$.
The solution of Eq.~\eqref{Eq:observables_not_closed} is, in general, a formidable task.
If one is interested in the properties of the steady-state, however, the time derivatives can be set to zero.
This approximation is  not sufficient to solve Eq.~\eqref{Eq:observables_not_closed}, since some operators are a function of higher-order
correlation functions, thus resulting in an infinite hierarchy of coupled equations.
In the normal phase (i.e., when no symmetry is broken), one can reduce the complexity of the problem by considering $\kappa \gg \Gamma_{\downarrow} \simeq  \Gamma_{\phi}$, i.e., that the bosonic field reaches a steady state long before the qubits do.
Indeed, in this region the Liouvillian gap must be opened \cite{MingantPRA18}, and in the absence of critical-slowing down the typical timescale is dictated by the dissipation rates.
Using adiabatic elimination, one can easily find the behavior of the system close to the normal phase characterised by $\moy{\hat{X}}=\moy{\hat{Y}}=\moy{\adag\aop}=\moy{\hat{J}_x}=\moy{\hat{J}_y}=0$, while $\moy{\hat{J}_z}=-1$.
The results of adiabatic elimination, however, fail to capture the superradiant phase: in this regime, the diverging timescale coming from the closure of the Liouvillian gap makes the photonic timescale comparable to the qubit one. 

In order to truncate the hierarchy of equations stemming from a Liouvillian problem, in many-body quantum physics one often resorts to a Gutzwiller mean-field approximation. 
In this case, one assumes that the system density matrix can be factorised as a tensor product between the qubit and photonic part, decoupling all the high-order correlation in Eq.~\eqref{Eq:observables_not_closed}.
For instance, we will assume that $\moy{\hat{J}_x \adag\aop}=\moy{\hat{J}_x} \moy{\adag\aop}$.
This mean field approximation is expected to be true for high-dimensional models (i.e., when the number of nearest neighbors is elevated) and in the thermodynamic limit $N \to \infty$.
This approximation was demonstrated to be valid also for the standard Dicke model in the thermodynamic limit \cite{Kirton17b, Shammah17}.
Under the mean field approximation, Eq.~\eqref{Eq:observables_not_closed} becomes
\begin{equation}\label{Eq:observables_closed}
    \begin{cases}
    \partial_t \moy{\hat{X}}=-\kappa \moy{\hat{X}}=-2i\omega_c \moy{\hat{Y}}, \\
    \partial_t \moy{\hat{Y}}=-\kappa \moy{\hat{Y}}-2i\omega_c \moy{\hat{X}} - 4ig\sqrt{N}\moy{\hat{J}_x} -8ig\sqrt{N}\moy{\hat{J}_x} \moy{\adag\aop},\\ 
    \partial_t \moy{\adag\aop}=2i g \sqrt{N}\moy{\hat{J}_x}\moy{\hat{Y}}-\kappa \moy{\adag\aop},\\
    \partial_t \moy{\hat{J}_x}=-2\omega_0\moy{\hat{J}_y} - \Gamma^\prime \moy{\hat{J}_x},\\
    \partial_t \moy{\hat{J}_y}=2\omega_0\moy{\hat{J}_x}-\Gamma^\prime \moy{\hat{J}_y}-\frac{2g}{\sqrt{N}}\moy{\hat{J}_z} \moy{\hat{X}},\\
    \partial_t \moy{\hat{J}_z}=\frac{2g}{\sqrt{N}}\moy{\hat{J}_y} \moy{\hat{X}}-\Gamma_{\downarrow}\moy{\hat{J}_z}-\Gamma_{\downarrow}.
    \end{cases}
\end{equation}
One solution to this equation is $\moy{\hat{X}}=\moy{\hat{Y}}=\moy{\adag\aop}=\moy{\hat{J}_x}=\moy{\hat{J}_y}=0$, while $\moy{\hat{J}_z}=-1$. That is, the normal phase with no photons is always a solution to Eq.~\eqref{Eq:observables_closed}.
However, there are other two solutions to this equation, where the bosonic field is populated. Namely:
\begin{equation}\label{Eq:solution}
    \begin{cases}
    \moy{\hat{X}}_{\rm ss}= \frac{8 g \omega_c \sqrt{N} \moy{\hat{J}_x}_{\rm ss}}{-(\kappa^2 + 4 \omega_c^2) + 16 g^2 N \moy{\hat{J}_x}_{\rm ss}^2}, \\
    \moy{\hat{Y}}_{\rm ss} = \frac{i \kappa}{2 \omega_c} \moy{\hat{X}}_{\rm ss}, \\
     \moy{\adag\aop}_{\rm ss}= \frac{2 i g \sqrt{N}}{\kappa}  \moy{\hat{J}_x}_{\rm ss} \moy{\hat{Y}}_{\rm ss},\\
    \moy{\hat{J}_x}_{\rm ss} = \pm \sqrt{\frac{\kappa^2 + 4 \omega_c^2}{16 g^2 N} + \frac{2 \omega_0 \omega_c}{N \left(4 \omega_0^2 + \Gamma'^2\right)} \moy{\hat{J}_z}_{\rm ss}},\\    
    \moy{\hat{J}_y}_{\rm ss} = - \frac{\Gamma'}{2 \omega_0}  \moy{\hat{J}_x}_{\rm ss},\\
    \moy{\hat{J}_z}_{\rm ss} = - \frac{1+ \beta}{2}+ \sqrt{\left(\frac{1+ \beta}{2}\right)^2 - \beta \frac{g_t^2}{g}},
    \end{cases}
\end{equation}
where we have introduced $\beta = \frac{\omega_c \Gamma'}{2 \omega_0 N \Gamma_\downarrow}$ and $g_t=\sqrt{\left(2 \omega_c + \frac{\kappa^2}{2 \omega_c} \right)\left(2 \omega_0 + \frac{\Gamma'^2}{2 \omega_0}\right)/8}$.

Having identified the three possible solutions, we study their stability by considering a linear perturbation of the steady-state value:
\begin{equation}
    \vec{A}=  \left\{ \moy{\hat{X}},\moy{\hat{Y}}, \moy{\adag\aop} , \moy{\hat{J}_x} ,  \moy{\hat{J}_y},  \moy{\hat{J}_z}\right\}^{\operatorname{T}} = \vec{A}_{\rm ss} + \delta\vec{A}.
\end{equation}
For the normal phase we have
\begin{equation}
    \partial_t \vec{A}=\partial_t(\delta \vec{A})=M_\text{N} \delta \vec{A} = \begin{bmatrix}-\kappa & -2i\omega_c & 0 & 0 & 0 & 0 \\ 
    -2i\omega_c & -\kappa & 0 & -4ig\sqrt{N} & 0 & 0 \\
    0 & 0 & -\kappa & 0 & 0 & 0  \\
    0 & 0 & 0 & -\Gp & -2\omega_0 & 0\\
    \frac{2g}{\sqrt{N}} & 0 & 0 & 2\omega_0 & -\Gp & 0 \\
    0 & 0 & 0 & 0 & 0 & -\Gamd \end{bmatrix} \delta \vec{A},
\end{equation}
while for the superradiant phase 
\begin{equation}
    \partial_t \vec{A}=\partial_t(\delta \vec{A})=M_\text{S} \delta \vec{A} =\begin{bmatrix}-\kappa & -2i\omega_c & 0 & 0 & 0 & 0 \\ 
    -2i\omega_c & -\kappa & -8ig\sqrt{N}\moy{\Jx}_s & -4ig\sqrt{N}(1+2\moy{\adag\aop}_s) & 0 & 0 \\
    0 & 2ig\sqrt{N}\moy{\Jx}_s & -\kappa & 2ig\sqrt{N}\moy{\Yop}_s & 0 & 0  \\
    0 & 0 & 0 & -\Gp & -2\omega_0 & 0\\
    -\frac{2g}{\sqrt{N}}\moy{\Jz}_s & 0 & 0 & 2\omega_0 & -\Gp &  -\frac{2g}{\sqrt{N}}\moy{\Xop}_s \\
    \frac{2g}{\sqrt{N}}\moy{\Jy}_s & 0 & 0 & 0 & \frac{2g}{\sqrt{N}}\moy{\Xop}_s & -\Gamd \end{bmatrix} \delta \vec{A}.
\end{equation}
Only if all the eigenvalues of $M_\text{N}$ ($M_\textbf{S}$) are negative, the normal (superradiant) phase is stable. We have obtained these eigenvalues numerically to study the phase diagram of the system.

\bibliographystyle{apsrev4-1}
\bibliography{References-updated.bib}

\section*{Acknowledgements}
\label{Acknowledgements}
N.S., S.F., and L.G. acknowledge hospitality by Marco Genoni and Matteo G.A. Paris in the Applied Quantum Mechanics group at the University of Milan, Italy. 
F.M. is supported by the FY2018 JSPS Postdoctoral Fellowship for Research
in Japan. F.N. acknowledges partial support from the MURI Center for
Dynamic Magneto-Optics via the Air Force Office of Scientific Research (AFOSR) award No. FA9550-14-1-0040, the
Army Research Office (ARO) under grant No. W911NF-18-
1-0358, the Asian Office of Aerospace Research and Development (AOARD) grant No. FA2386-18-1-4045, the Japan
Science and Technology Agency (JST) [through the Q-LEAP
program, the ImPACT program, and CREST Grant No. JPMJCR1676, the Japan Society for the Promotion of Science
(JSPS) through the JSPS-RFBR grant No. 17-52-50023 and
the JSPS-FWO grant No. VS.059.18N], the RIKEN-AIST
Challenge Research Fund, the FQXi and the NTT PHI Labs. S. F. acknowledges support from the European
Research Council (ERC-2016-STG-714870).

\section*{Author Information}
\label{auth}
\subsection*{Contributions}
All authors contributed to the manuscript. L.G. and P.W. studied the stability of the equations and contributed equally to this work. N.S. and S.F. conceived the work. N.S. and F.M. wrote the codes for the simulations. L.G. produced the plots. F.N. provided funding and supervised the project.

\end{document}